\documentclass[runningheads]{llncs}
\usepackage{graphicx}
\usepackage{pdfpages}
\usepackage{comment}

\usepackage{todonotes}
\usepackage{subcaption}

\usepackage{hyperref}
\hypersetup{
    colorlinks=true,
    breaklinks=true,
    urlcolor= blue,
    linkcolor= blue,
    bookmarksopen=false,
    filecolor=black,
    citecolor=blue,
    linkbordercolor=blue
}

\usepackage{url}

\usepackage{ amssymb }
\usepackage[misc,geometry]{ifsym}
\usepackage{orcidlink}

\usepackage{scalerel,xparse}

\begin{document}

\title{Visualization in the Era of Artificial Intelligence: Experiments for Creating Structural Visualizations by Prompting Large Language Models}
\titlerunning{Visualization in the Era of Artificial Intelligence}
%
\author{Hans-Georg Fill\inst{1}\orcidlink{0000-1111-2222-3333} \and Fabian Muff\inst{1}\orcidlink{1111-2222-3333-4444}}
\authorrunning{Fill and Muff}
%
\institute{University of Fribourg, Research Group Digitalization and Information Systems
\email{hans-georg.fill|fabian.muff@unifr.ch}\\
}

\maketitle            
\begin{abstract}
Large Language Models (LLMs) have revolutionized natural language processing by generating human-like text and images from textual input. However, their potential to generate complex 2D/3D visualizations has been largely unexplored. We report initial experiments showing that LLMs can generate 2D/3D visualizations that may be used for legal visualization. Further research is needed for complex 2D visualizations and 3D scenes. LLMs can become a powerful tool for many industries and applications, generating complex visualizations with minimal training.
\end{abstract}
\section{Introduction}

The creation and use of visualizations are common in many areas of science and practice. In particular, the field of legal visualization has a long historical tradition of using visualizations to explain complex legal relationships and scenarios~\cite{brunschwig2021visual}. The core advantage of visualization is that it provides a graphical representation of something that would otherwise be difficult for humans to understand. It is thus a form of complexity reduction, aiming at human comprehension and understanding. This can include large data sets that are easier for humans to understand in visual formats, and where certain properties of the data can only be discovered through visual means. Examples range from visual representations of business data, as commonly used in statistics and data analytics, to visualizations of medical data, e.g., to discover features in the human body, or visualizations of sensor data and their extrapolations to create animated weather maps~\cite{fill2009visualisation}. Visualization further includes visual representations of knowledge for facilitating human communication, cf.~\cite{eppler2004knowledge,fill2006semantic}. In contrast to visualizations of data that are typically generated through algorithms, such visualizations are created and curated by humans manually~\cite{fill2007technical}. Based on the intended purpose of the visualization, they may be based on a formal or semi-formal language – as typically used in conceptual modeling~\cite{hoppenbrouwers2005fundamental} – or may consist of ad-hoc compositions of shapes and drawings, e.g., as typically found in the area of infographics, cf.\cite{dunlap2016getting} or in structural legal visualization~\cite{cyras2015structural}.

The creation of this latter type of visualization typically requires not only a deep understanding of the domain and scenario to be represented but also skills for using appropriate tools and procedures ~\cite{brunschwig2021visual,fill2007technical}. These tools can range from drawing tools that allow the creation of arbitrary shapes or the reuse and adaptation of existing shapes, over modeling tools that require the use of predefined shapes and concepts, to sophisticated dedicated visualization tools, e.g., for the creation of 3D representations cf.~\cite{fill2015bridging}. While standard drawing tools are fairly easy to operate, modeling and specialized visualization tools often require significant training to use them effectively. 

With the recent emergence of highly capable large language models, this last aspect can be addressed in previously unimagined ways, especially through creating computer code in various languages~\cite{lehman2022evolution}. At their core, large language models are trained on very large amounts of text so that they represent a probabilistic model of word sequences~\cite{jurafsky_d_2023speech}. Each word in the vocabulary of the language model is thus assigned a probability that can follow a given input sequence or an already generated sequence of words. Recent scientific and technological advances have made it possible to increase the training sets for such language models to hundreds of billions of parameters~\cite{vaswani_a_2017attention}. In November 2022, the company OpenAI made one large language model publicly accessible in the form of a dialogue system called ChatGPT and a corresponding API. The underlying GPT language models have been trained on more than 175 billion parameters and are not only able to generate text sequences in different styles based on input sequences in the form of prompts but can also output code of programming languages and may soon be able to process multimodal data as well, e.g. images. 

Recently, the application of these large language models has been explored in a variety of domains to see how they can support the creation of text and code in various ways – e.g.,~\cite{fill2023conceptual}. Therefore, in the following, we report on some experiments we have conducted to create visualizations using prompts to a large language model in the form of GPT-4 . The goal of these experiments was to explore which types of visualizations may be generated in this way, with the particular goal of supporting at some point structural legal visualizations as used in legal informatics. Thereby we aim to contribute to the discussion on how novel types of artificial intelligence can support visualizations from a meta-perspective~\cite{sreevalsan2023metavisualization}. The approach we use in the following targets the generation of visualizations based on graphical primitives, i.e., using vector graphics or programming code. This is in contrast to other approaches of using large language models, for generating pixel graphics as done for example by Dall-E.

We structure the experiments according to the following categories: a) two-dimensional visualizations, b) simple three-dimensional visualizations for a web-based environment, and c) complex three-dimensional visualizations for dedicated 3D modeling editors. Thereby, the first category aims at structural legal visualizations as typically generated today manually, e.g., in PowerPoint, whereas the second and the third category target real three-dimensional visualizations, which are today being generated for example using programming libraries such as three.js  or the open-source 3D modeling toolkit Blender . The latter requires advanced knowledge of either programming or specialized 3D modeling software. Generating visualizations for these environments without this knowledge seemed to us the most interesting use case. 

\section{Prompts for Generating Two-Dimensional Visualizations}
The prompts we will show in the following were preceded by several trials to find out which type of instructions would lead to the most promising results. Further, it was required to assess which type of code for describing visualizations could be generated by GPT-4. Thereby we found that SVG (Scalable Vector Graphics) code, which is an open standard for representing vector graphics, could be successfully generated by GPT-4. Thus, the first prompt we show was formulated as follows:

\vspace{.3cm}
\noindent\fbox{\parbox{.97\textwidth}{Prompt 1: Create the visualization of two trees that are connected by a black line in SVG. Only show the code surrounded by triple backticks, do not add any explanation.}}
\vspace{.3cm}

\begin{figure}[]
	\begin{minipage}[t]{.24\textwidth}
        \centering
		\includegraphics[width=.93\textwidth]{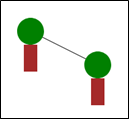}
		\subcaption{Prompt 1}
		\label{fig:1}
	\end{minipage}
	\begin{minipage}[t]{.24\textwidth}
        \centering
		\includegraphics[width=.89\textwidth]{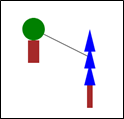}
		\subcaption{Prompt 2}
		\label{fig:2}
	\end{minipage}
 \begin{minipage}[t]{.24\textwidth}
        \centering
		\includegraphics[width=.96\textwidth]{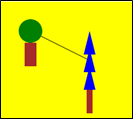}
		\subcaption{Prompt 3}
		\label{fig:3}
	\end{minipage}
 \begin{minipage}[t]{.24\textwidth}
        \centering
		\includegraphics[width=.93\textwidth]{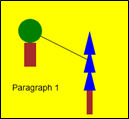}
		\subcaption{Prompt 4}
		\label{fig:4}
	\end{minipage}
    \caption{Results of Prompts one to four.}
	\label{fig:1to4}
\end{figure}

The result of this prompt is shown in Figure~\ref{fig:1}. We decided not to show the actual code generated by GPT4, but rather the result in the form of the resulting graphical representation when the code is executed in an SVG viewer application. The last sentence of the prompt refers to the display of the code result in the ChatGPT application, which can be forced to display the code in a separate area with this command. The example shows that the LLM can interpret the command and come up with a valid solution.  The type of prompt we used here is a so-called zero-shot prompt, which means that we rely on information already known to the large language model, i.e., it already knows about SVG code and how to create it. Nevertheless, it's fascinating that it can easily produce an abstract representation of a tree in this format. In the next experiment we stayed in the same session of the prompting and asked to modify the previous result with the following prompt:

\vspace{.3cm}
\noindent\fbox{\parbox{.97\textwidth}{Prompt 2: Change one tree to represent a fur with blue branches.}}
\vspace{.3cm}

The result can be seen in Figure~\ref{fig:2}. It correctly modified one tree and now shows a fur with blue branches. The next alteration we wanted to conduct was to change the surrounding of the trees. Thus we again stayed in the same session and executed this prompt:

\vspace{.3cm}
\noindent\fbox{\parbox{.97\textwidth}{Prompt 3: Put the two trees in the middle of a yellow lake.}}
\vspace{.3cm}

As shown in Figure~\ref{fig:3}, the result was not quite what we intended. The entire background of the image was now colored yellow. Of course, one could regard this as a valid result, but we would have imagined something more like a yellow ellipse surrounding the trees.
Finally, as common for many structural legal visualizations, we tried to add text to the previously generated image by the following prompt:

\vspace{.3cm}
\noindent\fbox{\parbox{.97\textwidth}{Prompt 4: Add the text "Paragraph 1" below the left tree.}}
\vspace{.3cm}

This worked equally well, as can be seen from Figure~\ref{fig:4}. The results we received via these prompts for generating two-dimensional visualizations are already quite promising. 

\section{Prompts for Generating Three-dimensional Visualizations}
Thus, we advanced by trying to generate three-dimensional representations as well. We found that GPT-4 is capable of generating valid code in the JavaScript programming language. This language is today very common in web-based environments and can be executed by all major web browsers. It also contains commands to access the WebGL API, which allows the creation of three-dimensional visualizations whose rendering can be accelerated directly by graphic processing units. Therefore, we have provided GPT-4 with the following prompt:

\vspace{.3cm}
\noindent\fbox{\parbox{.97\textwidth}{Prompt 5: Generate JavaScript code for displaying a simple house with yellow walls and a red roof using WebGL. Surround the code by triple backticks and do not add any explanation.}}
\vspace{.3cm}

\begin{figure}[]
	\begin{minipage}[t]{.3\textwidth}
  \centering
		\includegraphics[width=.6\textwidth]{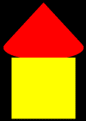}
		\subcaption{Prompt 5}
		\label{fig:5}
	\end{minipage}
	\begin{minipage}[t]{.33\textwidth}
  \centering
		\includegraphics[width=.84\textwidth]{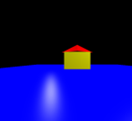}
		\subcaption{Prompt 6}
		\label{fig:6}
	\end{minipage}
 \begin{minipage}[t]{.34\textwidth}
  \centering
		\includegraphics[width=.97\textwidth]{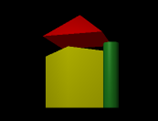}
		\subcaption{Prompt 7}
		\label{fig:7}
	\end{minipage}
    \caption{Results of Prompts five to seven.}
	\label{fig:5to7}
\end{figure}

We then executed the resulting code in a standard Firefox web browser and took a screenshot, shown in Figure~\ref{fig:5}. Despite the three-dimensional representation in the code, the visualization shown only resembles a 2D image. We found that the mechanisms for interacting with the three-dimensional representation were missing. This led to the next experiment as shown in Prompt 6. Here, we reverted to the three.js API, an open-source library for abstracting from the core WebGL code and making it easier to describe three-dimensional scenes. GPT-4 is also capable of generating code for this API.

\vspace{.3cm}
\noindent\fbox{\parbox{.97\textwidth}{Prompt 6: Create the code for displaying a three-dimensional house on a website using Three.js. The house shall have yellow walls and a red roof. It stands on a blue lake, which is made of a material that reflects objects. Add several lights to the scene. Only display the code using triple backticks without any explanations.}}
\vspace{.3cm}

The result of Prompt 6 is shown in Figure~\ref{fig:6}. Now, we can already better grasp the three-dimensional nature of the scene. Although three.js does have mechanisms for interacting with scenes, this has not been automatically added. So in Prompt 7, we specifically asked to include an animation loop so that the scene would be animated. Further, we asked for adding several lights to the scene. Although this prompt still does not require programming knowledge, it requires familiarity with some concepts used by three.js for displaying three-dimensional scenes.

\vspace{.3cm}
\noindent\fbox{\parbox{.97\textwidth}{Prompt 7: Create the code for displaying a three-dimensional house on a website using Three.js. The house shall have yellow walls and a red roof. It stands on a blue lake, which is made of a shiny material that reflects objects. In front of the house there is a fur tree. Add several lights to the scene. Add an animation loop to the scene so that the camera flies in a circle around the house. Only display the code using triple backticks without any explanations.}}
\vspace{.3cm}

The result of Prompt 7 is shown in Figure~\ref{fig:7}. Now, the user sees an animated version of the scene where the camera rotates around the house. Unfortunately, the requested blue lake, which had been correctly included in previous iterations, was now missing from this result as well as a correct tree representation.

Although three.js is a very advanced API for displaying 3D scenes, it always has to be embedded in the context of a website, which requires additional code to be generated. Due to current limitations of the accessible versions of ChatGPT, the amount of code that can be generated is rather limited. Therefore, we were interested in trying a more efficient approach. The open-source tool Blender is a full-fledged three-dimensional modeling and animation tool. It can be used to generate any kind of three-dimensional representation, including videos of animations. It is freely available, well-documented, and greatly supported by a large open-source community. In addition, Blender provides a scripting interface based on the Python programming language. This permits to access almost any functionality of the tool programmatically, i.e., by writing programming code and executing it, instead of accessing the user interface. One typical use case for this scripting interface is the creation of multiple 3D objects in particular arrangements, which would require a lot of manual effort. For example, this approach would allow a script to create hundreds of random three-dimensional objects without a single click. For our purposes, we discovered that GPT-4 can generate valid Python code based on the Blender libraries, which permits the generation of 3D scenes from natural language descriptions. As shown in Prompt 8 we executed the following instruction:

\vspace{.3cm}
\noindent\fbox{\parbox{.97\textwidth}{Prompt 8: Create the Python code for Blender for the following scene: There is a house with a red roof and white walls with blue windows. The camera looks at the house slightly from above and a point light points to the top of the roof. The house is placed on a large white plane where shadows are casted by the point light. Only show the code surrounded by triple backticks and do not add any explanation.}}
\vspace{.3cm}

\begin{figure}[b!]
    \centering
    \includegraphics[width=\textwidth]{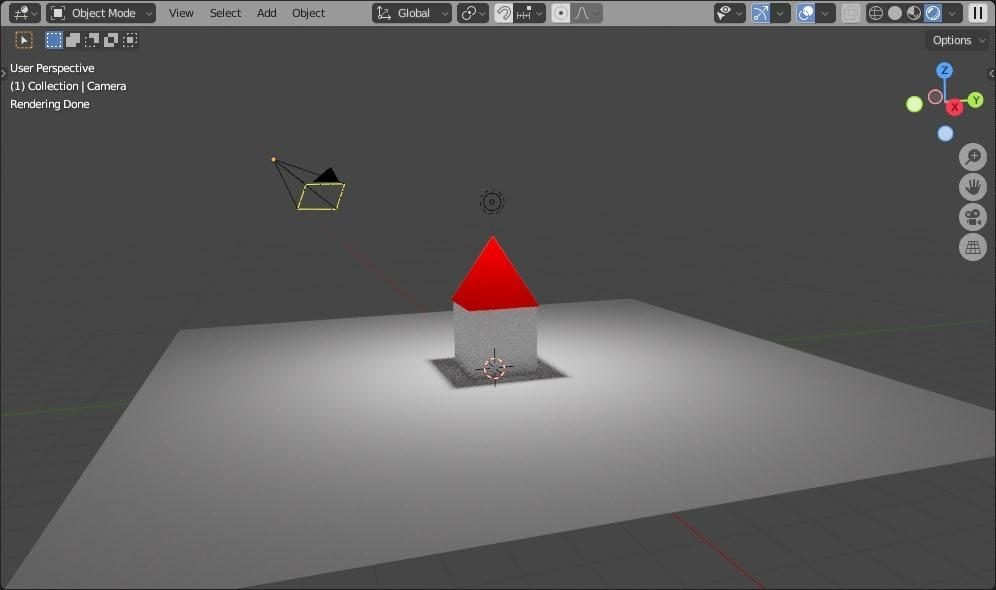}
    \caption{Scene generated by the Python scripting interface in Blender using the result of Prompt 8}
    \label{fig:8}
\end{figure}

This prompt generated Python code that could be directly executed using the scripting interface of Blender. The result of the Blender user interface is shown in Figure~\ref{fig:8}. It correctly displays a three-dimensional scene including a camera and a light, all with the correct positioning. From there, this scene could be directly further extended or modified, or exported in a large variety of file formats, e.g., to integrate it in a PowerPoint presentation. Although a user would still need to have the skills to do this in Blender, the main step of creating the 3D objects did not require any knowledge of 3D modeling at all, but was based on natural language statements.

Finally, we issued a more complex prompt for generating Python code for Blender. As shown in Prompt 9, we added further objects to the scene and wanted to see how GPT-4 would represent a car with yellow windows. 

\vspace{.3cm}
\noindent\fbox{\parbox{.97\textwidth}{Prompt 9: Create the Python code for Blender for adding the following items to the scene above: A tree with a brown trunk and green leaves stands in front of the house. A blue car with yellow windows stands next to the house. An arrow points from the car to the tree. Only show the code surrounded by triple backticks and do not add any explanations.}}
\vspace{.3cm}

\begin{figure}
    \centering
    \includegraphics[width=\textwidth]{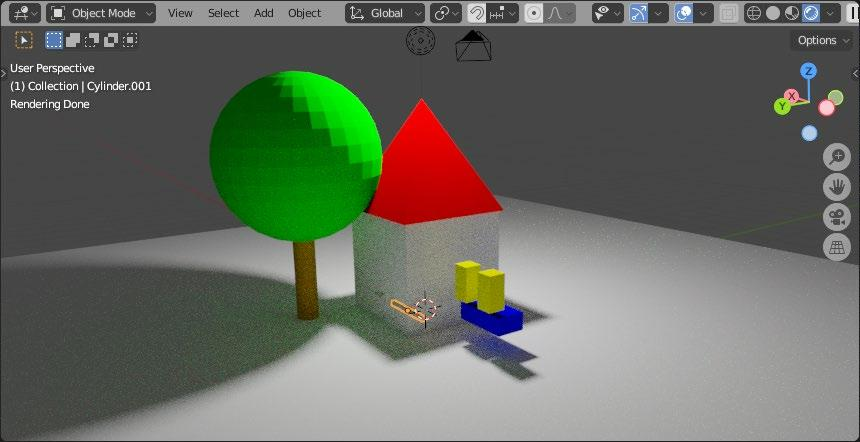}
    \caption{Result of the Blender scene created by Prompt 9}
    \label{fig:9}
\end{figure}

The result of this prompt is shown in Figure~\ref{fig:9}. The tree is nicely modeled using a green sphere and a brown cylinder representing the trunk of the tree. However, the car is very abstract and is not recognizable, as is the arrow that we tried to add.

\section{Discussion}
The experiments permit several interesting insights into the current and future possibilities of using large language models for generating structural visualizations. First, none of the experiments required us to write a single line of code. Rather, the code was only generated and then pasted into a corresponding execution environment (i.e., an SVG viewer, the Browser, and the Blender Python scripting editor). It can be expected that the further advancement of large language models will offer direct integration into such execution environments, e.g., via plugins for Blender that directly access the GPT API, which would already be technically feasible today. Then, there would be a seamless transition from specifying scenes in natural language and viewing them in the corresponding tool. In terms of the quality of the generated visualizations, we only showed prompts for comparatively simple scenarios. We have also tried to describe more complex structural legal visualizations, such as those available on Prof. Friedrich Lachmayer's website at legalvisualization.com, but have not yet been able to assemble complex scenes based on a textual description. It turned out that textual descriptions of visualizations that are precise enough for code generation become quite extensive, especially for positioning objects to each other in three-dimensional space. In such cases, it is still much easier to arrange objects using the mouse. However, for creating 3D objects that can then be arranged, the use of large language models seems an interesting option for the future as it permits also users not experienced in the details of 3D modeling to create objects which can then be positioned or scaled as needed for example. Unlike existing databases of 3D objects, users can create their 3D objects and even textures in their own style without knowing the technical details of the creation environment. For the two-dimensional models, the use of large language models worked rather well. Although we still need to explore further, which types of objects can be created in this way, the created visualizations already showed a high degree of exactness and matched mostly very well the issued prompts.

\section{Conclusion and Outlook}
In this short paper, we reported on first experiments for creating structural visualizations via large language models. The insights we gained may serve as inspiration for further experiments by the legal visualization community. For example, in the future one could imagine trying to create comics of legal scenarios in this way, or maybe even animations for use in legal design. From our side, we plan to further investigate how large language models can be trained or instructed in a way to better understand the different visualization formats and integrate them into conceptual visual modeling platforms.

%
%
%

\bibliographystyle{splncs04}
\bibliography{bibliograpy}
\end{document}